\documentclass[a4paper]{article}

\usepackage{INTERSPEECH2021}
\usepackage{color,soul}
\usepackage{hyperref}
\usepackage{multirow, stfloats}

\title{Hi-Fi Multi-Speaker English TTS Dataset}
\name{Evelina Bakhturina, Vitaly Lavrukhin, Boris Ginsburg, Yang Zhang
\thanks{Preprint. Submitted to INTERSPEECH-21}
}

\address{NVIDIA, Santa Clara, USA}
\email{\{ebakhturina, vlavrukhin, bginsburg, yangzhang\}@nvidia.com}

\begin{document}

\maketitle
\begin{abstract}
This paper introduces a new multi-speaker English dataset for training text-to-speech models. The dataset is based on LibriVox audiobooks and Project Gutenberg texts, both in the public domain. The new dataset contains about 292 hours of speech from 10 speakers with at least 17 hours per speaker sampled at 44.1 kHz. To select speech samples with high quality, we considered audio recordings with a signal bandwidth of at least 13 kHz and a signal-to-noise ratio (SNR) of at least 32 dB. 
The dataset is publicly released at ``http://www.openslr.org/109/''.
\end{abstract}
 
\section{Introduction}

\begin{table*}[bp]
\caption{English text-to-speech datasets}
\label{tab:datasets_limits}
\centering
\begin{tabular}{ccccccc}
\toprule
  \multirow{2}{*}{Dataset}   &
  \multicolumn{1}{l}{Num of} &
  \multicolumn{1}{l}{Avg num of} & \multicolumn{1}{l}{Sampling}  & 
  \multirow{2}{*}{SNR analysis} & 
  \multirow{2}{*}{License} &
    \multirow{2}{*}{Purpose}
  \\
  & \multicolumn{1}{l}{speakers}  & 
  \multicolumn{1}{l}{hours/speaker} & \multicolumn{1}{l}{rate, kHz}  \\ 
\midrule
  LJSpeech & 1    & 24  & 22.05 & - & Public Domain&
  single-speaker TTS\\
  M-AILABS & 3    & 34  & 16 & - & Permissive & single- and multi-speaker TTS\\
  VCTK     & 109  & 0.4 & 48 & - & CC BY 4.0 & multi-speaker / adaptive TTS \\
  LibriTTS & 2456 & 4.2 & 24 & Y & CC BY 4.0 &
  multi-speaker TTS\\
    Blizzard-2013 & 1 & 319 & 44.1 & professional speaker & Non-commercial & 
  single-speaker TTS\\
\midrule
\textbf{Hi-Fi TTS}& 10 & 29.2 & 44.1 & Y & CC BY 4.0 & high-quality multi-speaker TTS\\
\bottomrule
\end{tabular}
\end{table*}

Neural network (NN) based text-to-speech (TTS) systems are capable of synthesizing speech that sounds very close to natural speech~\cite{prenger2018waveglow, yang2020multiband, kong2020hifigan}. To produce high-quality speech, TTS models require speech samples recorded in professional studio with usually a few dozens of hours per speaker. 
Public datasets for training TTS systems are mostly constructed from LibriVox audio books~\cite{librivox} and text from Project Gutenberg\footnote{\url{http://gutenberg.org/}}.
Among the most popular datasets are the LJSpeech~\cite{ljspeech17} and the M-AILABS~\cite{mailabs}. 
The LJSpeech is a single-speaker TTS dataset derived from LibriVox  books. The corpus contains about 24 hours of speech sampled at 22.05 kHz. 
The M-AILABS is a multi-language corpus also based on LibriVox. The English subset of the M-AILABS contains about 102 hours of audio data for 3 speakers (2 female and 1 male). 
These two datasets are frequently used for single-speaker TTS research since they provide a relatively large number of hours per speaker~\cite{beliaev2020talknet,zeng2020melglow, kong2020hifi,fastpitch2021, chen2021adaspeech}.
However, low sampling rates and the lack of audio quality analysis limit the usability of the datasets for producing high-quality speech. 
The largest high-quality English  dataset for speech synthesis was provided by The Voice Factory as part of  the Blizzard Challenge 2013~\cite{BC2013}. The dataset has over 300 hours of  speech from a professional speaker, but this corpus is distributed under a non-commercial license.

Two highly popular datasets for multi-speaker TTS are the LibriTTS corpus~\cite{zen2019libritts} and voice cloning toolkit (VCTK) from CSTR~\cite{vctk2019}. The LibriTTS is a revised version of the LibriSpeech~\cite{panayotov2015librispeech} that excludes examples with low SNR. The LibriTTS is often used for research in low resource multi-speaker speech synthesis~\cite{park2019,valle2020} and voice adaptation~\cite{chen2021adaspeech}. 
The English multi-speaker part of VCTK contains short studio quality audio clips recorded by 109 native English speakers. Each speaker reads out about 400 sentences, which were selected from a newspaper, the rainbow passage and an elicitation paragraph used for the speech accent archive.  The total duration of the dataset is about 44 hours. All recordings were done with a sampling frequency of 96 kHz, and then down-sampled to 48 kHz. The dataset was intended for speaker-adaptive and multi-speaker TTS~\cite{ping2017deep, jia2018transfer, chen2020multispeech, chen2021adaspeech}. Both datasets gave a significant boost to the speech research thanks to large amount of data, wide speaker diversity, and a non-restrictive license. 
However, building high-quality TTS systems from such small amount of data per speaker is challenging.
Table \ref{tab:datasets_limits} summarizes the existing datasets. 

To accelerate the TTS research, we built a new dataset with the following goals in mind: 
\begin{itemize}
    \item speech signal's bandwidth of at least 13 kHz
    \item SNR of at least 32 dB in the 0.3 - 4 kHz band (where most speech energy is concentrated)
    \item high-quality reference texts, i.e. texts are normalized and text-audio match verified
    \item minimum sampling frequency of 44.1 kHz 
    \item minimum of 10 native English speakers 
    \item gender diversity 
    \item at least 15 hours of audio per speaker 
\end{itemize}

We fulfilled our goals by constructing the Hi-Fi Multi-Speaker English TTS (Hi-Fi TTS) dataset. 
The dataset includes speech data for 10 speakers (6 female and 4 male) with at least 17 hours per speaker (17.7-58.0 hours). The quality of the reference texts was checked by running inference with ASR models and including only samples with zero Word Error Rate (WER).

We expect the Hi-Fi TTS dataset to facilitate training of TTS models that 1) generalize better, i.e. have a broader range of pitch and fewer mispronunciations  2) have higher overall quality, i.e. less noisy and contain fewer artifacts, and 3) are more expressive and provide a wider range of voice variations to choose from.

Although the source material for the Hi-Fi TTS dataset is in the public domain, when we selected a voice sample for a text-to-speech (TTS) application we sought the written consent of the speaker. We believe it is ethically responsible to get the speaker's consent before using their voice for synthetic replication and we strongly encourage other developers to seek such consent.

The rest of the paper is organized as follows. Section 2 summarizes the process of speaker selection based on the audio quality, defines the data pre-processing steps and the text-audio alignment approach. Section 3 shows the final dataset statistics and available subsets.

\section{The dataset construction}

\subsection{Speaker selections} 
This section describes the process of selecting LibriVox readers with high-quality audio and sufficient amount of speech.

\subsubsection{Downloading metadata and audio samples}

First we download book and speaker metadata for all available English audio books from LibriVox.org. 
We use LibriLight~\cite{librilight}, LibriVox API\footnote{\url{https://librivox.org/api/info}} and internal scripts for parsing LibriVox book pages to get the required information about books and speakers and verify that the book links to the reference texts are valid.
We compute the total number of hours available per speaker and keep only readers with at least 50 hours of audio. 
Next, we download the most recent audio samples for the selected readers. Note, each book page on Librivox.org contains links to both individual chapters and the entire book in ZIP format. The ZIP archive contains audio files in MP3 format at a sampling rate of 22.05 kHz, while the individual chapter links redirect to 128 kbps MP3 audio files sampled at 44.1 kHz. We use the latter option to download 44.1 kHz audio files. At the end of this stage we obtain the most recent audio recordings for 250 readers with at least 50 hours of audio data and valid reference book links.

\begin{figure*}[t]
  \centering
  \includegraphics[width=\linewidth]{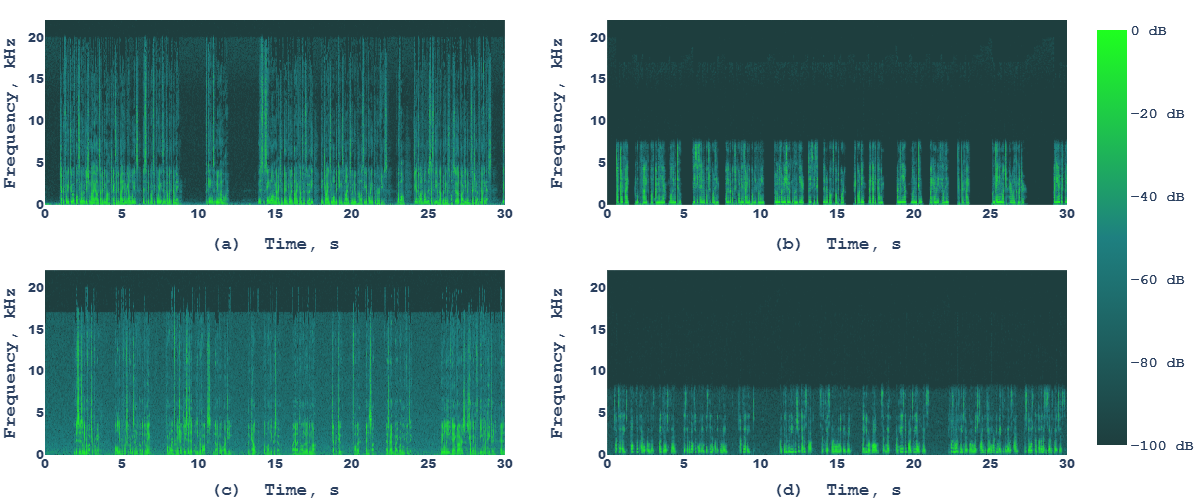}
  \caption{Spectrograms of LibriVox recordings: (a) good audio quality (wideband signal, low noise); (b) not acceptable audio quality (narrowband signal); (c) not acceptable audio quality (low SNR); (d) not acceptable audio quality (both narrowband and noisy).}
  \label{fig:speaker_selection}
\end{figure*}

\subsubsection{Audio quality analysis} \label{aq_analysis}
To select speakers with high audio quality, we estimate bandwidth and SNR to guide the selection process. We use the first 30 seconds of the audio samples for analysis.

To estimate the speech signal's bandwidth, we take the mean of the power spectrogram and find the highest frequency that has at least -50 dB level relative to the peak value of the spectrum.

\begin{figure}[h]
  \centering
  \includegraphics[width=\linewidth]{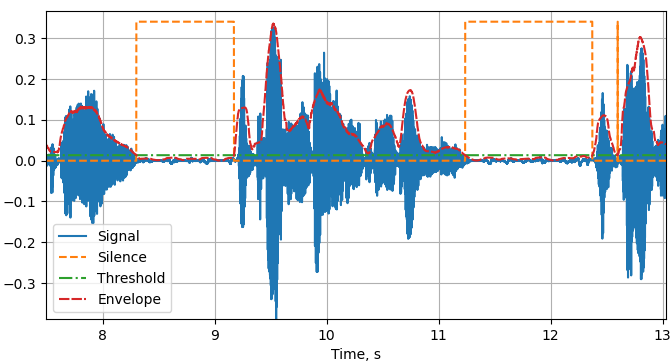}
  \caption{Energy-based VAD applied to a LibriVox recording.}
  \label{fig:vad}
\end{figure}

To estimate SNR we 1) pass the signal through a simple energy-based voice activity detector (VAD), 2) estimate the noise power in non-speech segments, 3) gauge the signal-plus-noise power in speech segments, and finally 4) compute SNR. This algorithm assumes that the noise is stationary. Energy-based VAD is demonstrated in Figure \ref{fig:vad}. This approach works fairly well in scenarios where SNR is not very low.

We consider four frequency bands for our SNR estimation. The frequency band from 100 Hz to 1 kHz helps to detect low-frequency tones and noises. The 300 Hz to 4 kHz band contains most of the speech energy, and we require at least 32 dB SNR in this band. Finally, the bands from 4 kHz to 10 kHz and from 10 kHz to 15 kHz are useful to spot high-frequency noises and to check if the signal's bandwidth is indeed wide.

The LibriVox readers were sorted based on SNR calculated for the most recently added book, and only the readers with the highest SNR values are considered. We assume that the recordings that were added last have the highest quality for the particular speaker.

We observe that many samples have a narrow signal bandwidth despite having a sampling rate of 44.1 kHz (see Figure \ref{fig:speaker_selection}.d). Further, in some cases non-speech segments of audio samples were denoised so that the assumption that noise is stationary was violated (see Figure \ref{fig:speaker_selection}.b). To verify that the selected speakers are a good match for the Hi-Fi TTS dataset, we listened to audio samples from each reader. 

After selecting an initial set of readers with acceptable audio quality, we calculate SNR for all the books of the selected readers since we observed  that the reader's audio quality may differ across the books. Based on SNR we select  books that have low noise and high bandwidth into the \textit{clean} subset of the dataset. The books with noticeable noise or narrower bandwidth are included in the \textit{other} subset (see Section \ref{data_stats}). We anticipate that some research project could benefit from additional data for readers even though the audio quality is not as high.
The SNR-based filtering along with the careful analysis of spectrogram reduces the number of the readers from 250 to 10 with 3 readers in the \textit{clean} set, and 7 readers in the \textit{other} set, see Table~\ref{tab:dataset_stats}. 

\begin{table*}[ht]
\caption{Hi-Fi Multi-Speaker TTS Dataset statistics.}
\label{tab:dataset_stats}
\centering
\scalebox{1}{
\begin{tabular}{@{}clcccc@{}}
\toprule
\textbf{Reader ID} & \textbf{Reader Name} & \textbf{Gender}      & \textbf{Clean, hrs} & \textbf{Other, hrs} & \textbf{Total, hrs} \\ \midrule
 92  & Cori Samuel     & F & 27.3 & -    & 27.3  \\
6097 & Phil Benson     & M & 30.1 & 3.4  & 33.5  \\
9017 & John Van Stan   & M & 58.0 & -    & 58.0  \\
6670 & Mike Pelton     & M & -    & 17.7 & 17.7  \\
6671 & Tony Oliva      & M & -    & 24.1 & 24.1  \\
8051 & Maria Kasper    & F & - & 30.4  & 30.4  \\
9136 & Helen Taylor    & F & -    & 24.3 & 24.3  \\
11614  & Sylviamb      & F & -    & 22.2 & 22.2  \\
11697 & Celine Major    & F & -   & 26.8 & 26.8  \\
12787 & LikeManyWaters  & F & -   & 27.3 & 27.3  \\ \midrule
\multicolumn{1}{l}{}  & \multicolumn{1}{l}{} & \multicolumn{1}{l}{} & \textbf{115.4}       & \textbf{176.2}       & \textbf{291.6}\\ \bottomrule
\end{tabular}}
\end{table*}

\subsection{Text pre-processing}

The text pre-processing pipeline can be summarized as follows:
\begin{enumerate}
    \item Download reference text files. In most cases, a LibriVox book page contains a link to the corresponding reference book in TXT format hosted at the Gutenberg Project. Rarely, LibriVox redirects to the reference book in PDF format, in such cases, we use the output of an Optical Character Recognition (OCR) from the PDF of the book. 
    \item Split reference text files into chapters based on the  LibriVox audio files. The original book text is a single file while the audio book is usually divided into chapters. To segment the original text into chapters that correspond to the chapter level audio recordings, we transcribe audio files with the QuartzNet ASR model~\cite{kriman2020quartznet}. Then, we match the chapter level audio with the corresponding text snippet using the transcripts along with the chapter titles information from LibriVox pages.
    \item Remove footnotes or inaudible sections in square or curly brackets from the reference texts. 
    \item Normalize text. All abbreviations and non-standard words~\cite{taylor_tts} are expanded to their full spoken equivalent with the publicly available NeMo normalization tool~\cite{zhang2021nemo}. For example, \textit{"on the 18th"} is converted to \textit{"on the eighteenth"}, \textit{"Hon."} to \textit{"Honorable"}, etc.
    \item Separate long chapter-level reference texts into chunks based on the end of sentence (EOS) punctuation marks to ensure that sufficient amount of samples represent complete sentences. As discussed in Zen et al.~\cite{zen2019libritts}, audio samples split on sentence breaks should facilitate learning long-term characteristics of speech such as sentence-level prosody features. The threshold of 60 characters was chosen empirically to ensure the average duration of the final segments is no longer than 20 seconds. The text chunks longer than 60 characters are further divided with semi-colon, dash or colon punctuation marks. 
\end{enumerate}

\subsection{Audio segmentation and text-audio alignment}

Next, we split long LibriVox audio files of the selected high-quality readers into smaller fragments suitable for TTS model training and align them with the corresponding reference texts.
The length of LibriVox audio files varies from a few minutes to an hour. At the same time, TTS models are trained on short audio clips that are usually up to 20 seconds long. To segment long audio files into short audio segments and the corresponding text fragments, we follow the work of K{\"u}rzinger et al. on Connectionist Temporal Classification (CTC)~\cite{graves2006connectionist} based segmentation ~\cite{ctcsegmentation}.\footnote{\url{https://pypi.org/project/ctc-segmentation/}}
 

LibriVox audio files contain preamble segments where readers briefly talk about the LibriVox project and state their name and book information. The preamble text is missing from the reference texts. To handle such cases, the CTC-Segmentation~\cite{ctcsegmentation} algorithm allows the alignment to start at any point of the audio file by setting the transition cost for the beginning of the segment to zero. However, when an utterance is mentioned in the untranscribed preamble and then repeated again as part of the actual recording the alignment results could be negatively affected. We observe that cutting off 3 seconds from the beginning of the audio helps to resolve this issue.

The CTC-Segmentation \cite{ctcsegmentation} consists of  two stages:
\begin{enumerate}
    \item Forward pass. Character log probabilities for each time step are obtained from an ASR model pre-trained with CTC loss. For this step we downsample the original 44.1 kHz MP3 audio files to 16 kHz and convert them to WAV format. Then we run the audio files through the QuartzNet\footnote{\url{https://ngc.nvidia.com/catalog/models/nvidia:nemospeechmodels}} ASR model~\cite{kriman2020quartznet} to extract the log probabilities (greedy decoding strategy without language model). The QuartzNet model was trained on six datasets: LibriSpeech~\cite{panayotov2015librispeech}, Mozilla Common Voice~\cite{ardila2019common}, the Wall Street Journal dataset~\cite{paul1992design}, The Fisher corpus \cite{cieri2004fisher}, Switchboard~\cite{godfrey1992switchboard}, and the Singapore English subset of the National Speech Corpus~\cite{koh2019building}. 
    \item Backward pass. At this stage, the most probable time step that corresponds to the last character of the text segment is determined and then the full path is recovered.
\end{enumerate}
Along with the split intervals the CTC-Segmentation technique emits a confidence score for each aligned segment. This score represents the probability of finding the correct alignment of the short text segment within the audio clip. The alignment confidence score goes down when the words in the reference text and audio clips do not match, for example when a reader flips, misses, or repeats words. We apply a threshold score value of $-2$ to remove poorly aligned utterances (the score value lies in log space).

\subsection{Post-processing and sample selection} \label{post_processing}

After aligning the text-audio pairs, we observe that a mismatch between audio transcript and reference text is still present, i.e. merely applying filtering based on the confidence alignment score is not sufficient. This is usually the case when the reference book version is changed or when a reader modifies the reference text while reading. To filter out such cases we run inference with both the QuartzNet and Citrinet \cite{citrinet} ASR models (greedy decoding, no language model) to calculate WER. Only samples with zero WER are included in the dataset. 
\section{Dataset splits and statistics} \label{data_stats}

\begin{figure}[t]
  \centering
  \includegraphics[width=\linewidth]{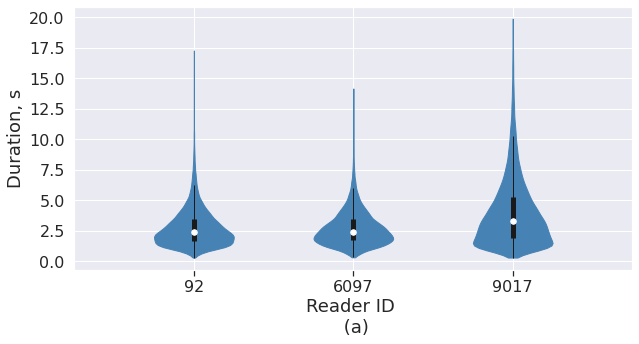}
  \includegraphics[width=\linewidth]{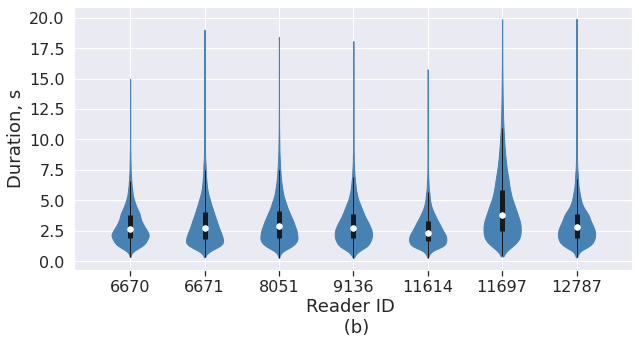}
  \caption{Violin plots of the audio lengths per speaker: \\ a) clean set, b) other set.}
  \label{fig:duration}
\end{figure}

The Hi-Fi TTS dataset contains audio clips sampled at 44.1 kHz in WAV format for 10 speakers (6 female and 4 male speakers), see(Table \ref{tab:dataset_stats}). Each speaker has at least 17 hours of speech. All  samples were verified and have zero WER (see Section \ref{post_processing}).
The Hi-Fi TTS corpus contains two parts:
\begin{itemize}
    \item The \textit{clean} subset contains high-quality audio books with relatively good voice attributes based on the SNR analysis (at least 40 dB), see Section \ref{aq_analysis}. 
    \item The \textit{other} set includes books with lower SNR (at least 32 dB). Some readers are present in both subsets as the audio quality varies from book to book.
\end{itemize}
Figure \ref{fig:duration} demonstrates violin plots~\cite{hintze1998violin} of the duration of the audio clips in the Hi-Fi TTS for each speaker.

We also created \textit{dev} and \textit{test} splits to facilitate model comparison. We provide original text and the corresponding normalized version of the text (with letter capitalization preserved). 
At the time of writing, the Hi-Fi TTS does not overlap with the M-AILABS and the LJSpeech datasets. The LibriTTS most likely contains some audio samples from speakers included in the Hi-Fi TTS.

\section{Conclusions}

 This paper introduces the Hi-Fi Multi-Speaker TTS dataset created from the LibriVox audiobooks and the Gutenberg project texts. The readers included in the dataset were carefully selected using SNR and spectrogram analysis. The quality of the text-audio alignments and the accuracy of the reference texts were validated with ASR model predictions (only samples with zero WER are accepted). To our knowledge, this is the first multi-speaker dataset of such quality. 
 
 Despite the careful selection of the speakers and their books, the audio quality of LibriVox recordings is still lagging far behind professional audio recordings. The next step of the project could be to apply voice processing techniques to improve the quality of the sound, for example by removing hissing and metallic sound in samples with boosted high frequencies.
 
 The code for normalization, segmentation and interactive data analysis is publicly available via NeMo toolkit \cite{kuchaiev2019nemo}.\footnote{\url{https://github.com/NVIDIA/NeMo}}

\section{Acknowledgements}
The authors would like to thank Rafael Valle, Adrian Lancucki, Jocelyn Huang and the rest of the NVIDIA TTS team for review, helpful comments and feedback. We also thank the LibriVox volunteers without whom this project would not have been possible.

\bibliographystyle{IEEEtran}
\bibliography{mybib}
\end{document}